# Piezoelectricity in Two-Dimensional Group III Monochalcogenides

Wenbin Li[1] and Ju Li[2,1]

[1]Department of Materials Science and Engineering, Massachusetts Institute of Technology, Cambridge, MA 02139, United States

[2]Department of Nuclear Science and Engineering, Massachusetts Institute of Technology, Cambridge, MA 02139, United States

**Abstract**

We find that several layer-phase group-III monochalcogenides, including GaS, GaSe and InSe, are piezoelectric in the monolayer form. First-principles calculations reveal that the piezoelectric coefficients of monolayer GaS, GaSe and InSe are on the same order of magnitude as the earlier discovered two-dimensional piezoelectric materials, such as BN and $MoS_2$ monolayers. Our study expands the family of two dimensional piezoelectric materials, suggesting that strong piezoelectric response can occur in a wide range of two dimensional materials with broken inversion symmetry. The co-existence of piezoelectricity and superior photo-sensitivity in these two-dimensional semiconductors enables the integration of electromechanical and optical sensors on the same material platform.

**Main Text**

Piezoelectric materials have a wide range of applications in systems that require robust electrical-mechanical coupling, including mechanical stress sensors, actuators and energy harvesting devices[6-11]. Technology progress continues to push for increasingly miniaturized devices, motivating search of piezoelectric materials with nanoscale dimensions. For a crystal to be piezoelectric, it must not possess inversion symmetry[12]. Many layer-phase materials from which atomic-thick two-dimensional (2D) materials[13] are derived have inversion symmetry in the bulk. However, when thinned down to monolayer, inversion symmetry may no longer be present and the monolayer has the potential to be piezoelectric. Recently, several 2D materials, specifically boron nitride (BN) monolayer and some transition metal dichalcogenide (TCMD) monolayers were theoretically predicted to be intrinsically piezoelectric with large piezoelectric coefficients[5], which have been experimentally demonstrated for molybdenum disulfide ($MoS_2$) monolayer[10,11]. The high crystal quality and large elastic strain limits[14,15] of such 2D materials could enable high performance piezoelectric materials at nanoscale[10,11]. It is expected that there are more 2D piezoelectric materials than hitherto discovered.

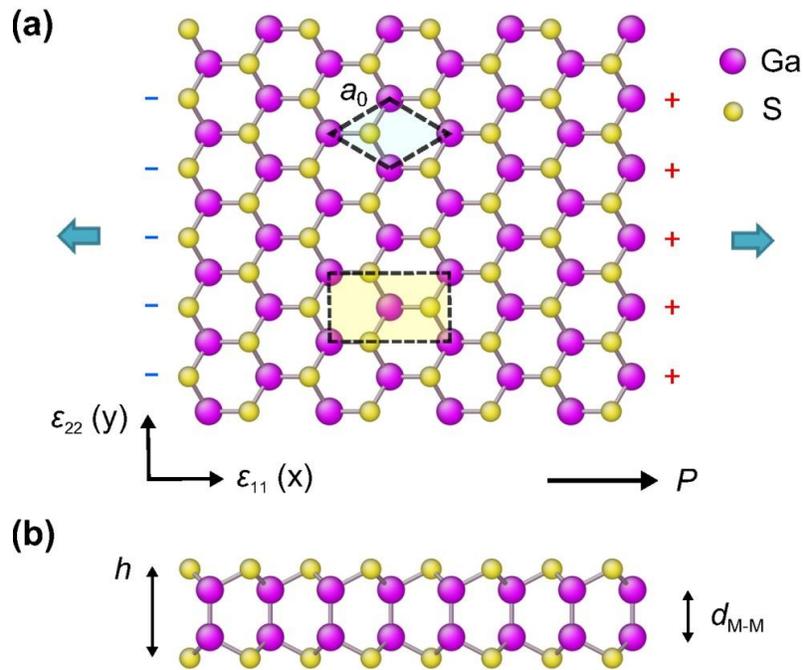

**Figure 1.** Structural model of gallium sulfide (GaS) monolayer viewing from (a) top and (b) side, which is representative of the structure of all three MX monolayers in this study. The larger-sized gallium atoms are purple, while the sulfur atoms are yellow. In (a), a two-dimensional primitive cell with rhombus shape and lattice constant $a_0$ is highlighted. A rectangular unit cell drawn beneath is used for DFT calculations. The x direction of the orthogonal coordinate system corresponds to the "armchair" direction of the monolayer, while the y direction corresponds to the "zigzag" direction. The direction of piezoelectric polarization after uniaxially stretching the monolayer along the x direction is labeled. (b) Side view of the GaS monolayer. The thickness of monolayer, as defined by the distance between the top and bottom layer of sulfur atoms, is denoted by $h$. Also indicated is the distance between two bonded metal atoms, $d_{M-M}$.

Characterization of new 2D piezoelectric materials not only expands the family of piezoelectric nanomaterials, but potentially also enables new phenomena and functionalities through cross-coupling between electrical, chemical, optical and mechanical responses in novel material systems.

In this Letter, we use first-principles density functional theory (DFT) calculations to demonstrate that several monolayer group-III monochalcogenides (MX, M = Ga or In, X = S or Se), including gallium sulfide (GaS), gallium selenide (GaSe) and indium selenide (InSe), are intrinsically piezoelectric with large piezoelectric coefficients. The bulk crystals of GaS, GaSe and InSe belong to the same GaSe-type structure, consisting of vertically stacked X-M-M-X layers held together by weak van der Waals-like forces. Within each four-atom-thick monolayer the bonding is mainly covalent with some ionic

**Table 1. Experimental and DFT-PBE calculated structural parameters and bandgap for bulk and monolayer MX.** The values of monolayer lattice constant $a_0$, monolayer thickness $h$, bond length between metal atoms $d_{M-M}$ and bandgap $E_{gap}$ are listed.

| material | bulk experiment[1-4] | | | | monolayer DFT-PBE calculation | | | |
|---|---|---|---|---|---|---|---|---|
| | $a_0$ (Å) | $h$ (Å) | $d_{M-M}$ (Å) | $E_{gap}$ (eV) | $a_0$ (Å) | $h$ (Å) | $d_{M-M}$ (Å) | $E_{gap}$ (eV) |
| GaS | 3.59 | 4.60 | 2.45 | 3.05[a] | 3.64 | 4.66 | 2.48 | 2.37 |
| GaSe | 3.76 | 4.78 | 2.46 | 2.02[b] | 3.82 | 4.83 | 2.47 | 1.81 |
| InSe | 4.00 | 5.28 | 2.78 | 1.17[c] | 4.09 | 5.39 | 2.82 | 1.40 |

[a,b,c] Bandgap values measured at $T$ = 77 K, 300 K and 300 K respectively. Data from ref.[4].

character. Depending on the stacking sequence of the monolayers, several different structural polytypes exist, which are named $\beta$, $\varepsilon$, $\gamma$ and $\delta$ phase respectively[2]. In the past, these layered semiconductors attracted considerable attention for their nonlinear optical properties[16]. Interest in MX was revived recently after the discovery that many layered compounds can be exfoliated down to monolayers with relative ease[13]. Study of 2D MX is currently an active field of research, and monolayers or few-layers of all the three compounds studied in paper have either been mechanically exfoliated[17-22] or chemically synthesized *via* vapor phase transport[23,24], exhibiting excellent potential for use in flexible photodetectors[18,19,21].

Figure 1 shows the top and side views of the atomic structure of the GaS monolayer, representative of the monolayer structure of all three MX compounds. Two vertically bonded atomic layers of metal atoms are sandwiched between two layers of chalcogen atoms. Viewings from the top, the M and X atoms occupy one triangular sub-lattice of honeycomb lattice respectively. The structure of MX monolayer thus bears strong similarity to the monolayers of transition metal dichalcogenide (TMDC) of 2H type, such as $MoS_2$. In fact, both MX and $MoS_2$ monolayers belong to $D_{3h}(\bar{6}m2)$ point group symmetry. A notable feature is the absence of inversion center in the symmetry group, leading to potential non-zero piezoelectric response of MX monolayers, as in the case of BN and 2H-TMDC monolayers[5,10,11]. We thus carried out first-principles calculations to quantify the piezoelectric properties of MX monolayers.

Our DFT calculations were carried out using the Vienna Ab initio Simulation Package (VASP) with a plane wave basis set[25,26] and the projector-augmented wave method[27]. Exchange-correlation functionals in the Perdew-Burke-Ernzerhof (PBE) form[28] within the generalized gradient approximation (GGA)[29] were used. A plane-wave cutoff energy of

**Table 2.** DFT-PBE calculated in-plane elastic stiffness $C_{11}$ and $C_{12}$ of monolayer MX. The Poisson ratio $v_\perp$ is also calculated for the relaxed-ion case.

| material | clamped-ion | | relaxed-ion | | |
|---|---|---|---|---|---|
| | $C_{11}$ (N/m) | $C_{12}$ (N/m) | $C_{11}$ (N/m) | $C_{12}$ (N/m) | $v_\perp$ |
| GaS | 108 | 32 | 83 | 18 | 0.39 |
| GaSe | 91 | 26 | 70 | 16 | 0.34 |
| InSe | 75 | 35 | 51 | 12 | 0.40 |

450 eV was used throughout all calculations. To facilitate the calculation of unit cell polarization, we used an orthorhombic unit cell containing four M atoms and four X atoms, as indicated in Figure 1a. The cell height in the direction normal to the monolayer plane (z direction) was 20 Å. Brillouin zone integration employed an 11 × 18 × 1 Γ centered Monkhorst-Pack[30] k-point grid, corresponding to k-point sampling spacing less than 0.1 Å$^{-1}$ along the reciprocal lattice vectors in the x-y plane. Convergence criteria for electronic and ionic relaxations were 10$^{-6}$ eV and 10$^{-3}$ eV/Å respectively.

The DFT-optimized structure parameters, including lattice constant $a_0$, monolayer thickness $h$ and M-M bond length $d_{\text{M}-\text{M}}$, are listed in Table 1. We also include in Table 1 the corresponding experimental structural parameters in the bulk phase[1-3], allowing direct comparison. The structural parameters from our DFT calculations are in perfect agreement with a previous study[31]. The slightly larger values of the structural parameters compared to the bulk experimental values is attributed to the use of PBE exchange-correlation functional, which tends to overestimate lattice parameters by one to two percent[32]. As these materials are semiconductors, the experimental bandgap values for the bulk phase and the DFT calculated values for monolayers are also listed in Table 1. Calculations indicated that all three MX compounds are indirect bandgap semiconductors in the monolayer limit. The calculated bandgap values, however, are for reference only as ground-state DFT calculations tend to significantly underestimate the value of bandgap[33]. More accurate bandgap values of monolayers through excited states calculations at $G_0W_0$[33,34] level and hybrid functional calculations have been calculated by Zhuang et al.[31].

To facilitate comparison with previous calculation results of 2D piezoelectric materials, our calculation of the piezoelectric coefficients of MX monolayers followed procedures very similar to Duerloo et al.[5]. We first obtained the planar elastic stiffness coefficients

$C_{11}$ and $C_{12}$ of the MX monolayers by fitting the DFT-calculated unit-cell energy $U$ to a series of 2D strain states ($\varepsilon_{11},\varepsilon_{22}$), based on

$$C_{11} = \frac{1}{A_0}\frac{\partial^2 U}{\partial \varepsilon_{11}^2}, \quad C_{22} = \frac{1}{A_0}\frac{\partial^2 U}{\partial \varepsilon_{22}^2} \text{ and } C_{12} = \frac{1}{A_0}\frac{\partial^2 U}{\partial \varepsilon_{11}\varepsilon_{22}}, \quad (1)$$

where $A_0$ is the unit-cell area at zero strain. Since $C_{11} = C_{22}$ for the MX monolayers with $D_{3h}$ point group symmetry, in the small strain limit we can write

$$\Delta u(\varepsilon_{11}, \varepsilon_{22}) = \frac{1}{2}C_{11}(\varepsilon_{11}^2 + \varepsilon_{22}^2) + C_{12}\varepsilon_{11}\varepsilon_{22}, \quad (2)$$

where $\Delta u(\varepsilon_{11}, \varepsilon_{22}) = [U(\varepsilon_{11}, \varepsilon_{22}) - U(\varepsilon_{11} = 0, \varepsilon_{22} = 0)]/A_0$ is the change of unit-cell energy per area. We carried out strain energy calculation on a 7×7 grid with $\varepsilon_{11}$ and $\varepsilon_{22}$ ranging from -0.006 to 0.006 in steps of 0.002. The atomic positions in the strained unit cell were allowed to fully relax. $C_{11}$ and $C_{12}$ based on the relaxed atomic coordinates are named relaxed-ion stiffness coefficients, which are experimentally relevant. If the atomic positions are not allowed to relax after applying unit-cell strain, then the so-called clamped-ion coefficients, which are of theoretical interest, can be calculated. Table 2 summarizes the clamped- and relaxed-ion stiffness coefficients for the three MX monolayers. In the relaxed-ion case, the Poisson ratios $\nu_\perp$ are also given, obtained directly from relaxed ion coordinates by evaluating the change of layer thickness in response to in-plane hydrostatic strain $\Delta h/h = -\nu_\perp(\varepsilon_{11} + \varepsilon_{22})$. The values of relaxed-ion $C_{11}$ coefficients show good agreement with earlier calculation[31].

We next calculate the linear piezoelectric coefficients of the MX monolayers by evaluating the change of unit-cell polarization after imposing uniaxial strain on the system, based on the modern theory of polarization[35,36] as implemented in VASP. The linear piezoelectric coefficients $e_{ijk}$ and $d_{ijk}$ are third-rank tensors as they relate polarization vector $P_i$, to strain $\varepsilon_{jk}$ and stress $\sigma_{jk}$ respectively, which are second-rank tensors:

$$e_{ijk} = \frac{\partial P_i}{\partial \varepsilon_{jk}}, \quad (3)$$

$$d_{ijk} = \frac{\partial P_i}{\partial \sigma_{jk}}. \quad (4)$$

Since our focus is on quasi-2D systems, the $i, j, k$ indices can be 1 or 2, corresponding to *x* or *y* directions, as indicated on Figure 1. The number of non-zero and unique piezoelectric coefficients is restricted by symmetry elements of the crystal[12]. For 2D monolayers with $D_{3h}$ point group symmetry, the symmetry properties of piezoelectric coefficients has been analyzed by Duerloo *et al.*[5], where it was concluded that the only unique piezoelectric coefficients are $e_{111}$ and $d_{111}$. Written in Voigt notation as $e_{11}$ and $d_{11}$, they are further related through the elastic stiffness coefficients as

$$e_{11} = d_{11}(C_{11} - C_{12}). \tag{5}$$

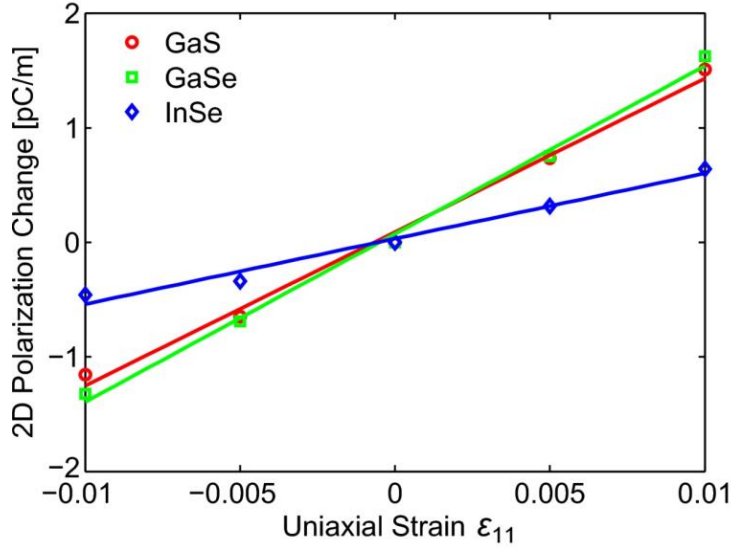

**Figure 2.** Change of unit-cell polarization per area of the MX monolayers along the *x*-direction after applying uniaxial strain $\varepsilon_{11}$ along the same direction. Ionic positions within the unit cells were relaxed after imposing strain to the unit cell. The piezoelectric coefficients $e_{11}$ correspond to the slope of lines obtained through linear fitting of polarization change with respect to $\varepsilon_{11}$.

We have directly calculated the piezoelectric coefficients $e_{11}$ of the MX monolayers by applying uniaxial strain $\varepsilon_{11}$ to the orthorhombic unit cell along the *x* direction, and then evaluating the differential change of unit-cell polarization per area along the *x* direction. The values of $e_{11}$ reported here resulted from the linear fitting of 2D polarization change $\Delta P_1$ with respect to $\varepsilon_{11}$ for $\varepsilon_{11}$ ranging from $-0.01$ to $0.01$ at intervals of 0.005, which is illustrated in Figure 2. Similar to the evaluation of elastic stiffness coefficients, clamped-ion and relaxed-ion coefficients can be obtained respectively depending on whether the ionic positions are allowed to relax after applying strain. The relaxed-ion (or clamped-ion) $d_{11}$ coefficients are then calculated from corresponding $e_{11}$ coefficients and elastic stiffness coefficients $C_{11}$ and $C_{12}$ based on Equation 5.

Table 3 lists the calculated clamped-ion and relaxed-ion piezoelectric coefficients $e_{11}$ and $d_{11}$ of the MX monolayers. The corresponding values of monolayer BN and $MoS_2$ calculated by Duerloo et al.[5], as well as the experimental measured value of $e_{11}$ for $MoS_2$ monolayer[11], are also included for comparison. We have benchmarked our calculation for $MoS_2$ monolayer, reaching good agreement with previous calculation results[5] (within 15% of difference due to the use of different simulation packages and pseudopotentials, etc.). As can be seen from Table 3, the calculated relaxed-ion piezoelectric coefficients of the GaS and GaSe monolayers ($e_{11}$ = 1.34 × 10$^{-10}$ C/m, $d_{11}$

**Table 3.** Calculated clamped-ion and relaxed-ion piezoelectric electric coefficients, $e_{11}$ and $d_{11}$, of GaS, GaSe and InSe monolayers. The piezoelectric coefficient of h-BN and MoS$_2$ monolayers calculated by Duerloo et al.[5], as well as the experimental value of $e_{11}$ for MoS$_2$ monolayer[11], are listed for comparison.

| material | clamped-ion | | relaxed-ion | |
|---|---|---|---|---|
| | $e_{11}$ ($10^{-10}$ C/m) | $d_{11}$ (pm/V) | $e_{11}$ ($10^{-10}$ C/m) | $d_{11}$ (pm/V) |
| GaS | 5.39 | 8.29 | 1.34 | 2.06 |
| GaSe | 5.22 | 9.67 | 1.47 | 2.30 |
| InSe | 5.17 | 13.26 | 0.57 | 1.46 |
| h-BN cal.[5] | 3.71 | 1.50 | 1.38 | 0.60 |
| MoS$_2$ cal.[5] | 3.06 | 2.91 | 3.64 | 3.73 |
| MoS$_2$ exp.[11] | | | 2.9 | |

= 2.06 pm/V and $e_{11}$ = 1.47 × $10^{-10}$ C/m, $d_{11}$ = 2.30 pm/V, respectively), are on the same order of magnitude as MoS$_2$ monolayer ($e_{11}$ = 3.64 × $10^{-10}$ C/m, $d_{11}$ = 3.73 pm/V), and in general larger than those of BN monolayer ($e_{11}$ = 1.38 × $10^{-10}$ C/m, $d_{11}$ = 0.60 pm/V). The piezoelectric coefficients of InSe monolayer ($e_{11}$ = 0.57 × $10^{-10}$ C/m, $d_{11}$ = 1.46 pm/V) are smaller than GaS and GaSe monolayers but still comparable to BN monolayer. In contrary to MoS$_2$ monolayer, relaxing ion positions after applying strain significantly reduces the polarization of the MX monolayer, a trend consistent with BN monolayer. As a result, the clamped-ion piezoelectric coefficients of the MX monolayers are much larger than the relaxed-ion coefficients.

We note that, previous experimental studies have demonstrated that the piezoelectric coefficients of MoS$_2$ monolayer shows non-monotonic decrease with the number of layers, resulting from the breaking and recovery of inversion symmetry in odd and even number of layers[10,11]. This effect shall also exist in the MX monolayers, due to the similarity of layer-stacking sequence between bulk MX and MoS$_2$.

From an application perspective, experimental studies have demonstrated that MX monolayers or few-layers have far superior photo-responsivity and faster response time than MoS$_2$ monolayers[18,19,21]. Although the piezoelectric coefficients of MX monolayers are slightly smaller than those of MoS$_2$ monolayers, the piezoelectric properties of MX monolayers can find applications where excellent photo-sensitivity and electro-mechanical coupling are both valued. For example, MX monolayers can be used to make flexible photodetectors where mechanical and optical signals can be electrically sensed based on the same active material.

In conclusion, we have computational demonstrated that several single-layer group-III monochalcogenides, specifically GaS, GaSe and InSe, are piezoelectric with linear piezoelectric coefficients on the same order of magnitude as other two-dimensional piezoelectric materials discovered previously. The good piezoelectric properties, combined with the excellent photo-sensitivity of these two-dimensional semiconductors, enable the integration of stress-sensor, photo-detectors and possibly nanoscale transducer on the same material platform, with potential applications in flexible electronics, optoelectronics and electromechanical systems.

We gratefully acknowledge financial support by NSF-DMR-1120901.